\documentstyle[multicol,prl,aps,psfig]{revtex}

\begin{document}
\draft
\title{Spectroscopic signatures of spin-charge separation in the quasi-one-dimensional organic conductor TTF-TCNQ}

\author{R. Claessen,$^1$ M. Sing,$^1$, U. Schwingenschl{\"o}gl,$^1$ P. Blaha$^2$, M. Dressel$^3$, and C.S. Jacobsen$^4$}

\address{$^1$Experimentalphysik II, Universit\"at Augsburg, D-86135 Augsburg, Germany}

\address{$^2$Institut f\"ur Physikalische und Theoretische Chemie, Technische
Universit\"at Wien, A-1060 Wien, Austria}

\address{$^3$1.~Physikalisches Institut, Universit{\"a}t Stuttgart, D-70550 Stuttgart, Germany}

\address{$^4$Department of Physics, Technical University of Denmark, DK-2800 Lyngby, Denmark}

\date{\today }
\maketitle

\begin{abstract}
The electronic structure of the quasi-one-dimensional organic conductor
TTF-TCNQ is studied by angle-resolved photoelectron spectroscopy
(ARPES). The experimental spectra reveal significant discrepancies to band theory.
We demonstrate that the measured dispersions can be consistently mapped onto the
one-dimensional Hubbard model at finite doping. This interpretation
is further supported by a remarkable transfer of spectral weight as
function of temperature. The ARPES data thus show spectroscopic signatures
of spin-charge separation on an energy scale of the conduction band width.
\\[1ex]
\end{abstract}


\begin{multicols}{2}
\narrowtext

The Fermi liquid (FL) concept is a central paradigm of solid state physics.
It describes the low-lying electronic excitations of metals in terms of
renormalized and only weakly interacting quasiparticles. Recently, various
materials have been discovered whose unusual
low-energy properties signal a breakdown of the quasiparticle
picture. The theoretically best established case of non-FL physics is
that of the interacting one-dimensional (1D) electron gas \cite{Voit95}.
Here the quasiparticle picture fails due to a decoupling of
charge and spin degrees of freedom, leading to the emergence of a new
generic many-body quantum state commonly referred to as Luttinger
liquid. Precise results have been obtained for its spectral properties
at very low excitation energies \cite{Voit95,Meden92}. Recently,
several theoretical studies
\cite{Penc96,Penc97,Carmelo00,Senechal00,Carmelo01}
have pointed out that in the case of strong {\it local} Coulomb
interaction signatures of spin-charge separation may also be observable
at {\it high} energies on the scale of the conduction band width.

Experimentally, the search for Luttinger liquid signatures
in quasi-1D metals has so far been directed at their
low-energy spectral behavior. For example, virtually all
quasi-1D metals studied by angle-resolved photoelectron spectroscopy
(ARPES) display a characteristic suppression of spectral weight near
the Fermi level \cite{Grioni00,Gweon01}. However, this finding
alone is not sufficient as manifestation of generic non-FL physics.
Therefore, an unambiguous and direct spectroscopic
identification of spin-charge separation in a 1D metal is still
lacking.

In this Letter we present ARPES results on the electronic structure of
the quasi-1D organic conductor TTF-TCNQ (tetrathiafulvalene tetracyanoquinodimethane)
which reveal significant discrepancies from a conventional band structure description.
We demonstrate that the TCNQ-related part of the experimental spectra can be
consistently mapped onto the 1D Hubbard model at finite doping and thus
reflects signatures of spin-charge decoupling. This interpretation is further supported by
the observation of a temperature-dependent transfer of spectral weight
over an energy scale of the band width.

\begin{figure}[htbp]
\begin{center}\mbox{}
\psfig{figure=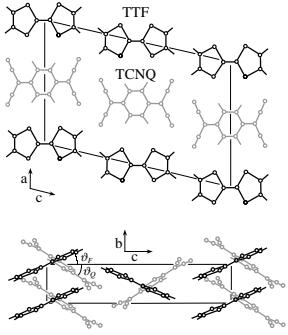,width=0.4\textwidth}
\end{center}
\caption{Crystal structure of TTF-TCNQ.}
\label{crystal structure}
\end{figure}

TTF-TCNQ is an organic charge transfer salt whose anisotropic properties
can be understood from its monoclinic crystal structure
\cite{Kistenmacher74,Kagoshima88}, depicted in Fig.~\ref{crystal structure}.
Planar TTF and TCNQ molecules form separate linear stacks along the
crystallographic {\bf b} direction. Molecular orbitals of
$\pi$-symmetry overlap with those of neighboring molecules stacked above and
below. The covalent bonding along {\bf b} is maximized by a rotation of
the molecular planes about the {\bf a}-axis.
A charge transfer of $\sim$0.59 electrons per
molecule from TTF to TCNQ drives both types of stacks metallic.
The conductivity along {\bf b} is up to three orders of magnitude
larger than perpendicular to it, making TTF-TCNQ a truly quasi-1D
metal. This is also reflected by a Peierls instability occuring at
$T_P = 54$ K.

There is substantial experimental evidence that Coulomb interaction plays an essential
role in the electronic structure of TTF-TCNQ \cite{Kagoshima88,Torrance77}.
The effect of strong electronic correlations can approximately be
described by the single-band 1D Hubbard model, with the on-site
({\it i.e.} intramolecular) Coulomb energy $U$ and the hopping
integral $t$ as parameters. Various properties of TTF-TCNQ have
successfully been analysed within this approach
\cite{Kagoshima88}. There is general agreement
that for TTF-TCNQ the ratio $U / 4t \sim 1$,
with the unperturbed bandwidth $4t\sim0.5$ eV from experiment and band
theory \cite{Kagoshima88,Torrance77}.

ARPES essentially measures the electron removal part of the spectral function
$A(k,\omega)$ \cite{95Huf}. The experiments were partly performed at the
U2/FSGM undulator beamline of BESSY I in Berlin, using an Omicron AR 65
spectrometer \cite{Janowitz99}. Spectra on the temperature dependence were
taken at our home lab using He I radiation from a rare gas
discharge lamp. In both cases the energy and momentum resolution
amounted to 60 meV and $0.07$ \AA$^{-1}$, respectively. All data
were taken above the Peierls transition. Clean surfaces parallel to the {\bf ab}-plane
were obtained by {\it in situ} cleavage of the
crystals at a base pressure of $< 10^{-10}$ mbar.
Surface quality was checked by x-ray photoemission \cite{Claessen01b}.
All data presented here were measured before
notable radiation-induced surface damage occurred.

\begin{figure}

\begin{center}\mbox{}
\psfig{figure=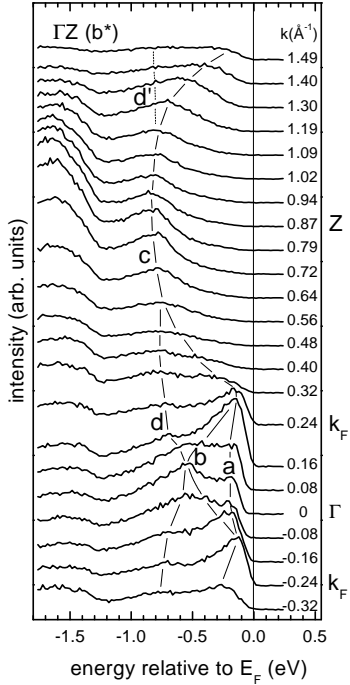,width=0.4\textwidth}
\end{center}
\caption{Angle-resolved photoemission spectra measured for wavevectors k along the
$\Gamma$Z direction ($h \nu = 25$ eV, $T = 61$ K). The thin lines indicate the
dispersion of the spectral features.}
\label{EDCs}
\end{figure}

The ARPES spectra measured along the ${\bf b}$ axis are shown in Fig.~\ref{EDCs} and
display pronounced dispersive behavior. Its existence and periodic symmetry
indicates a long-range surface order consistent with the bulk periodicity
along {\bf b}. Spectra measured
perpendicular to ${\bf b}$ are dispersionless (not shown here) as expected for a 1D metal.
Our data are in excellent agreement with those of Zwick {\it
et al.}\cite{Zwick98} but partly display more spectral detail.

We now turn to a detailed discussion of the dispersions in
Fig.~\ref{EDCs}. At zero momentum ($k = 0$) we can clearly identify two peaks at 0.19 and
0.54 eV below $E_F$. With increasing wavevector the intensity dip between them gets smeared out
leading to a broad intensity distribution between 0.15 and 0.6
eV, until at $k=0.16$ \AA$^{-1}$ again two peaks are resolved. This peculiar behavior is
consistent with a splitting of the 0.54 eV peak at $k=0$
into two separate structures, one moving to higher binding energy (labeled {\sf d} in
Fig.~\ref{EDCs}) and one ({\sf b}) dispersing towards the Fermi
level, until it eventually merges with the low binding energy peak {\sf a}
leading to the intense structure near the Fermi level at $k \approx 0.16...0.24$ \AA$^{-1}$.
For even higher momenta a weak structure
{\sf c} moves back again from the Fermi level and displays a dispersion
symmetric about $k=0.87$ \AA$^{-1}$, corresponding to the Brillouin zone edge
(Z-point \cite{criticalpoints}). Simultaneously, structure {\sf d}
disperses to higher binding energy and eventually becomes
obscured by peak {\sf c}. For wavevectors in the next zone
a symmetry-related weak shoulder {\sf d$^\prime$} is observed.

Close to $k=0.24$ \AA$^{-1}$ spectral features {\sf a},
{\sf b}, and {\sf c} reach their closest approach to the Fermi level.
We identify this position as Fermi vector, even though no actual
Fermi edge is observed as already noted by Zwick {\it et al.} \cite{Zwick98}.
Rather, the intensity is suppressed almost linearly right down to the
Fermi energy. We thus obtain a Fermi surface nesting vector of
$2k_F = 0.48 \pm 0.06$ \AA$^{-1}$, in good agreement with the
modulation vector of the Peierls state \cite{Kagoshima88}.

\begin{figure}
\begin{center}\mbox{}
\psfig{figure=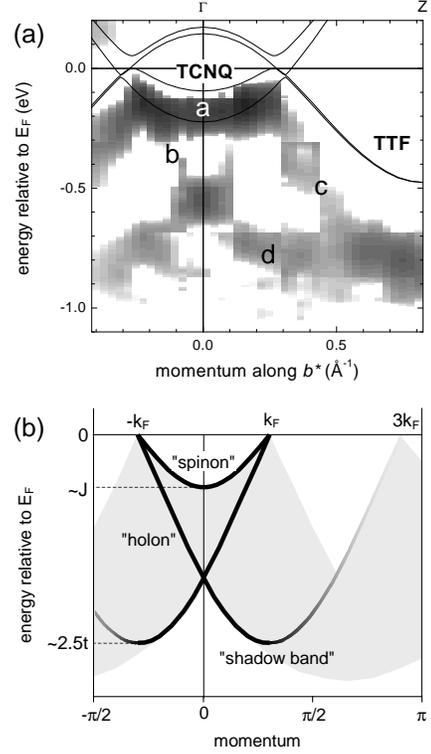,width=0.45\textwidth}
\end{center}
\caption{(a) Gray-scale plot of the ARPES dispersions (see text for details).
Also shown are the conduction bands obtained from density-functional band
theory. (b) Schematic electron removal spectrum of the 1D Hubbard model at
finite doping (band filling $n<2/3$).}
\label{intensity map}
\end{figure}

Fig.~\ref{intensity map}(a) displays the ARPES spectra as gray-scale plot of
their negative second energy derivative in the $(E,k)$-plane,
which enhances the visibility of the spectral structures and their
dispersion. Also shown is our density functional band calculation \cite{Blaha97a},
which yields two pairs of nearly degenerate
bands attributed to the TCNQ chains and TTF chains, respectively.
The {\it qualitative} behavior of experimental structures {\sf a}, {\sf b},
and {\sf c} is found to be in good correspondence with band theory.
Peaks {\sf a} and {\sf b} are thus attributed to the TCNQ chains,
while {\sf c} is assigned to the TTF stacks. {\it Quantitatively} we find however
that the experimental bandwidths exceed those of the calculation by
a factor of $\sim 2$. The fact that band theory is otherwise in
good agreement with bulk properties indicates that at the surface,
{\it i.e.} in the topmost molecular layers probed by ARPES, the
hopping integral $t$ and therefore the bandwidth is renormalized.
One may speculate that this effect is caused by a molecular surface relaxation;
it is however of little relevance for the following discussion.

A much more serious problem of band theory is its failure to
account for experimental feature {\sf d}. An interpretation
in terms of a surface state seems tempting, but is in conflict with
the observed Fermi vector: The additional charge
of a completely occupied surface state would severely affect the
delicate charge balance between the TTF and TCNQ bands and shift the
surface Fermi vector notably from its bulk value, which is
not observed. An alternative explanation of {\sf d} as {\it umklapp}
image of the TTF band induced by long-ranged Peierls fluctuations
\cite{Schaefer01} is ruled out due to the lack of other evidence of
backfolding in the data.

In view of the strong correlation effects encountered in other
properties of TTF-TCNQ we now compare the ARPES dispersions to the
electron removal spectra of the 1D Hubbard model at finite doping,
schematically depicted in Fig.~\ref{intensity map}(b). The (photo)hole
generated by removal of an electron decays into two collective
excitations with separate dynamics, a spinon characterized by a spin
quantum number and a spinless holon carrying the charge \cite{Voit95}.
As a consequence the electron removal spectrum consists of a broad
continuum determined by the phase space available for spinon-holon
decomposition, indicated by the gray-shaded region in
Fig.~\ref{intensity map}(b). However, dispersive singularities
may appear (solid lines in Fig.~\ref{intensity map}(b)) for certain
decompositions of the real hole. For example, the low binding energy
singularity arises from hole fractionalization into a holon bound to
the Fermi level and a propagating spinon; it therefore directly
reflects the dispersion of a bare spinon, whose bandwidth scales with
the exchange integral $J$. It is therefore referred to as "spinon"
branch. The "holon" branch corresponds to a spinon pinned to the Fermi
level and a propagating holon. It crosses the Fermi level at
$k_F \pm 2k_F$ \cite{Penc96,Carmelo01}. From its
energy minimum at $k_F$ up to $3k_F$
(referred to as "shadow band" in Ref.~\onlinecite{Penc96})
it looses strongly in intensity until it becomes barely distinguishable from the
diffuse background. Initially this spectral behavior was derived for
the strong coupling limit ($U >> 4t$) of the 1D Hubbard-model
\cite{Penc96,Favand97}. Very recently
it has been shown that it also holds for moderate interaction
strengths of $U \sim 4t$ as relevant for TTF-TCNQ
\cite{Senechal00,Carmelo01}.

Comparing Figs.~\ref{intensity map}(a) and (b) one finds a
surprisingly good agreement of the TCNQ-related ARPES dispersions
with the Hubbard model spectrum. The experimental structures
{\sf a} and {\sf b} closely resemble those of the theoretical
spin and charge branches, respectively \cite{footnote_on_splitting}.
Furthermore, the as yet unidentified structure {\sf d} can be well
accounted for as the "shadow band" of the Hubbard model,
at least for not too large $k$-vectors. Its upwards dispersion beyond
$k_F$ and its eventual $3k_F$-crossing are not observed, probably due
to the theoretically predicted loss of weight at larger $k$ and the
interfering TTF band. Within the Hubbard model interpretation the experimental
dispersions thus give direct spectroscopic evidence of spin-charge
separation in TTF-TCNQ.

\begin{figure}
\begin{center}\mbox{}
\psfig{figure=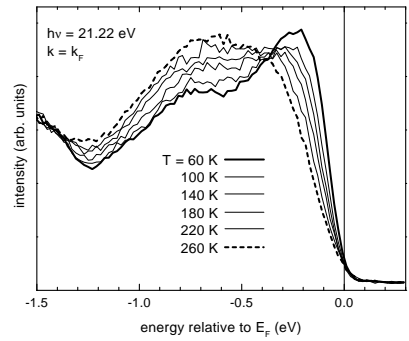,width=0.45\textwidth}
\end{center}
\caption{Temperature-dependence of the spectrum at $k=k_F$.
Note the dramatic transfer of weight from low to high binding energy with increasing temperature.}
\label{T-dependence}
\end{figure}

The Hubbard model interpretation of the experimental dispersions
is further supported by the temperature dependence of the ARPES
spectra measured at $k_F$, shown in Fig.~\ref{T-dependence}.
With increasing temperature from 60 to 260 K the spectra reveal
a dramatic shift of intensity from the low binding energy peak
({\sf a/ \sf b}) to structure {\sf d} at 0.75 eV. The effect is
fully reversible and conserves the total spectral weight within
experimental accuracy. The strong temperature dependence was already
reported in Ref.~\cite{Zwick98}, but the weight conservation had
been overlooked there due to a different intensity normalization
\cite{footnote_intensity_normalization}. A weight transfer over an energy scale
so much larger than $k_B T$ cannot be accounted for by conventional
electron-phonon coupling or the effect of Peierls fluctuations. On the other hand,
calculations of the temperature-dependent spectrum of the $tJ$ model at quarter filling
predict a redistribution of spectral weight on a scale of $t$ already for
temperatures $k_B T << t$ \cite{Penc97}, exactly as observed in our data.
A similar temperature dependent weight transfer has recently been
reported for ARPES spectra of the 1D Mott-Hubbard insulator
Na$_{0.96}$V$_2$O$_5$ \cite{Kobayashi99} but without a
resolution of individual "spinon" and "holon" peaks.

In summary, the unusual behavior of the
ARPES spectra of TTF-TCNQ is consistent with the 1D Hubbard model at finite doping,
thus reflecting spectral signatures of spin-charge separation over an
energy scale of the conduction band width.
We note however that the nearly linear spectral onset at {\it low} energies
is not reproduced by the simple Hubbard model.
Close to $E_F$ the density of states of an interacting
1D metal is  expected to follow a $|E-E_F|^\alpha$
power law behavior. With only on-site interactions the Hubbard model
yields $\alpha \leq 1/8$ \cite{Voit95}, at variance with the ARPES data.
Exponents of up to $\alpha \sim 1$ become possible for an extended Hubbard model including
nearest and next-nearest neighbor interactions \cite{Zhuravlev01}.
Based on the strong $2k_F$ correlations in TTF-TCNQ also the
Luther-Emery (LE) model has been suggested to provide an adequate description
of its low-energy spectrum \cite{Voit98}. Whether the extended Hubbard model or the LE model
are compatible with the observed {\it high}-energy behavior is presently unknown and
calls for further theoretical work.

We thank S.~Hao, Th.~Finteis, C.~Janowitz, and
G.~Reichardt for technical support at BESSY, and J.M.P.~Carmelo,
A.~Kampf, A.~Muramatsu, K.~Penc, and J.~Voit for
stimulating discussions. This work was supported by the
DFG (CL 124/3-1 and SFB 484) and
the BMBF (Grant 05SB8TSA2).


\begin{thebibliography}{10}

\bibitem{Voit95}
J. Voit, Rep.~Prog.~Phys. {\bf 58},  977  (1995).

\bibitem{Meden92}
V. Meden and K. Sch{\"o}nhammer, Phys. Rev. B {\bf 46},  15753  (1992).

\bibitem{Penc96}
K. Penc, K. Hallberg, F. Mila, and H. Shiba, Phys. Rev. Lett. {\bf 77},  1390
  (1996).

\bibitem{Penc97}
K. Penc, K. Hallberg, F. Mila, and H. Shiba, Phys. Rev. B {\bf 55},  15475
  (1997).

\bibitem{Carmelo00}
J.~M.~P. Carmelo, N.~M.~R. Peres, and P.~D. Sacramento, Phys. Rev. Lett. {\bf
  84},  4673  (2000).

\bibitem{Senechal00}
D. S\'{e}n\'{e}chal, D. Perez, and M. Pioro-Ladri{\`e}re, Phys. Rev. Lett. {\bf
  84},  522  (2000).

\bibitem{Carmelo01}
J.~M.~P. Carmelo, J.~M. B.~Lopos dos Santos, L.~M. Martelo, and P.~D.
  Sacramento, to be published.

\bibitem{Grioni00}
M. Grioni and J. Voit,  in {\em Electron Spectroscopies Applied to
  Low-Dimensional Materials}, edited by H. Starnberg and H. Hughes (Kluwer,
  Dordrecht, 2000), Vol.~1.

\bibitem{Gweon01}
G.-H. Gweon {\it et~al.}, J. Electron Spectrosc. Rel. Phen. {\bf 117-118},  481
   (2001).

\bibitem{Kistenmacher74}
T.~J. Kistenmacher, T.~E. Phillips, and D.~O. Cowan, Acta Cryst. {\bf B30},
  763  (1974).

\bibitem{Kagoshima88}
S. Kagoshima, H. Nagasawa, and T. Sambongi, {\em One-dimensional conductors}
  (Springer, Berlin, 1987), and references therein.

\bibitem{Torrance77}
J.~B. Torrance, Y. Tomkiewicz, and B.~D. Silverman, Phys. Rev. B {\bf 15},
  4738  (1977).

\bibitem{95Huf}
S. H{\"u}fner, {\em Photoemission Spectroscopy} (Springer, Berlin, 1995).

\bibitem{Janowitz99}
C. Janowitz {\it et~al.}, J. Electron Spectrosc. Rel. Phenom. {\bf 105},  43
  (1999).

\bibitem{Claessen01b}
R. Claessen {\it et~al.}, physica B, in print.

\bibitem{Zwick98}
F. Zwick {\it et~al.}, Phys. Rev. Lett. {\bf 81},  2974  (1998).

\bibitem{criticalpoints}
Our Z-point is identical to the Y-point of Ref.~\onlinecite{Zwick98}.

\bibitem{Blaha97a}
obtained with the \it WIEN97 \rm code (Vienna University of Technology 1997).
  [Improved version of P.~Blaha, K.~Schwarz, P.~Sorantin, and S.B.~Trickey,
  Comput.~Phys.~Commun.~{\bf 59}, 339 (1990)].

\bibitem{Schaefer01}
J. Sch{\"a}fer {\it et~al.}, Phys. Rev. Lett. {\bf 87},  196403  (2001).

\bibitem{Favand97}
J. Favand {\it et~al.}, Phys. Rev. B {\bf 55},  R4859  (1997).

\bibitem{footnote_on_splitting}
In this interpretation the doublet character of the TCNQ bands predicted by
  band theory is assumed to be unresolved in the ARPES spectra.

\bibitem{footnote_intensity_normalization}
Our spectra are normalized to incoming photon flux, which incidentally leads to
  an alignment of the spectra for binding energies $^{>}_{\sim} 1.3$ eV,
  independent of temperature.

\bibitem{Kobayashi99}
K. Kobayashi {\it et~al.}, Phys. Rev. Lett. {\bf 82},  803  (1999).

\bibitem{Zhuravlev01}
A.~K. Zhuravlev and M.~I. Katsnelson, Phys. Rev. B {\bf 64},  033102  (2001).

\bibitem{Voit98}
J. Voit, Eur.~Phys.~J.~B {\bf 5},  505  (1998).

\end{thebibliography}


\end{multicols}

\end{document}